\def\({ \left( }
\def\){ \right) }
\def\b{\begin{equation}}
\def\e{\end{equation}}
\def\={\ =\ }

\def\+{\ +\ }
\def\-{\ -\ }

\def\Ls{\cal L \rm}

\def\mumu{$\mu^+\mu^-$}
\def\ee{$e^+e^-$}

\documentstyle[epsfig]{aipproc} 
\renewcommand{\thesection}{\arabic{section}}
\renewcommand{\thesubsection}{\thesection.\arabic{subsection}}

\begin{document}
\title{MUON COLLIDERS }

\author{R.~B.~Palmer\thanks{Brookhaven National Laboratory,  Upton, NY 
11973-5000, USA}$^,$\thanks{Stanford Linear Accelerator Center, Stanford, CA 
94309, USA}, A.~Sessler\thanks{Lawrence Berkeley National Laboratory,
Berkeley, CA 94720, USA}, A.~Skrinsky\thanks{BINP, RU-630090 Novosibirsk,
Russia}, A.~Tollestrup\thanks{Fermi National Accelerator Laboratory,  
Batavia, IL 60510, USA},\\               
A.~J.~Baltz$^1$, P.~Chen$^2$, W-H.~Cheng$^3$, Y.~Cho\thanks{Argonne National
Laboratory, Argonne, IL 60439-4815, USA}, E.~Courant$^1$, R.~C.~Fernow$^1$, 
J.~C.~Gallardo$^1$,  A.~Garren$^{3,}$\thanks{ UCLA, Los Angeles, CA 90024-1547,
USA}, M.~Green$^{3}$, S.~Kahn$^1$, H.~Kirk$^1$, Y.~Y.~Lee$^1$, F.~Mills$^5$,  
N.~Mokhov$^{5}$, G.~Morgan$^1$,  D.~Neuffer$^{5,}$\thanks{ CEBAF, Newport News,
VA 23606, USA},  R.~Noble$^{5}$,  J.~Norem$^{6}$, M.~Popovic$^{5}$,  L.~Schachinger$^3$,
G.~Silvestrov$^4$, D.~Summers\thanks{ Univ. of Mississippi, Oxford, MS 38677,
USA}, I.~Stumer$^1$,  M.~Syphers$^1$,  Y.~Torun$^{1,}$\thanks{ SUNY, Stony
Brook, NY  11974, USA}, D.~Trbojevic$^1$,  W.~Turner$^{3}$,
A.~Van~Ginneken$^{5}$, T.~Vsevolozhskaya$^4$,  R.~Weggel\thanks{Francis Bitter
National Magnet Laboratory, MIT, Cambridge, MA 02139, USA}, E.~Willen$^1$,
D.~Winn\thanks{ Fairfield University, Fairfield, CT 06430-5195, USA},
J.~Wurtele\thanks{UC Berkeley, Berkeley, CA 94720-7300, USA} }
\maketitle
\thispagestyle{empty}
\begin{abstract}
Muon Colliders have unique technical and physics advantages and disadvantages
when compared with both hadron and electron machines. They should thus be
regarded as complementary. Parameters are given of 4 TeV and 0.5 TeV high
luminosity \mumu colliders, and of a 0.5 TeV lower luminosity demonstration
machine.
We discuss the various systems in such muon colliders,
starting from the proton accelerator needed to generate the muons and
proceeding through muon cooling, acceleration and storage in a collider ring.
Problems of detector background are also discussed.
\end{abstract}
\section{INTRODUCTION}
\subsection{Technical Considerations}
The possibility of muon colliders was introduced  by Skrinsky et
al.\cite{ref2}, Neuffer\cite{ref3},  and others. More recently, several
workshops and collaboration meetings have greatly increased the level of
discussion\cite{ref4},\cite{ref5}.  In this paper we discuss  scenarios for 4
TeV and 0.5 TeV  colliders based on an optimally designed proton source, and
for a lower luminosity 0.5 TeV demonstration based on an upgraded version of
the AGS. It is assumed that a demonstration version based on upgrades of the
FERMILAB machines would also be possible (see second Ref.\cite{ref5}).

Hadron collider energies are limited by their size and technical constraints on
bending magnetic fields. At very high energies it would also be difficult to
obtain the required luminosities, which must rise as the energy squared. \ee
colliders, because they undergo simple, single-particle interactions, can reach
higher energy final states than an equivalent hadron machine. However,
extension of $e^+e^-$  colliders to multi-TeV energies is severely
performance-constrained by beamstrahlung, and cost-constrained because two full
energy linacs are required\cite{ref1} to avoid the excessive synchrotron
radiation that would occur in rings. Muons (${m_{\mu}\over m_e}=207$) have the
same advantage in energy reach as electrons, but have negligible beamstrahlung,
and can be accelerated and stored in rings, making the possibility of high
energy \mumu colliders attractive. There are several major technical problems with
$\mu$'s:

\begin{itemize}
\item they decay with a lifetime of  $2.2\times 10^{-6}$ s. This problem is
partially overcome by rapidly increasing the energy of the muons, and thus
benefitting from their
relativistic $\gamma$ factor. At 2 TeV, for example, their lifetime is
$0.044\,$s: sufficient for approximately 1000 storage-ring collisions;
 \item another consequence of the muon decays
is that the decay products heat the magnets of the collider
ring and  create backgrounds in the detector;

\item  they are created through pion decay into a diffuse phase space and
this phase space cannot be reduced by conventional stochastic or synchrotron
cooling. It can, to some extent, be dealt with by ionization cooling, but
the final emittance of the muon beams will remain larger than that possible in
an \ee collider.
\end{itemize}

Despite these problems it appears possible that high energy muon
colliders might have luminosities comparable to or higher than those in \ee
colliders at the same energy\cite{ref6}. And because the \mumu machines would
be much smaller\cite{ref7}, and require much lower precision (the final
geometric emittances are about 5 orders of magnitude larger, and the spots are
about three orders of magnitude larger), they may be significantly less
expensive. It must be remembered, however, that a \mumu collider remains a new
and untried concept, and its study has just began; it cannot yet
be compared with the more mature designs for an \ee collider.
\subsection{Overview of Components}
\begin{figure}[t!] 
\caption{Schematic of a Muon Collider.}
\centerline{\epsfig{file=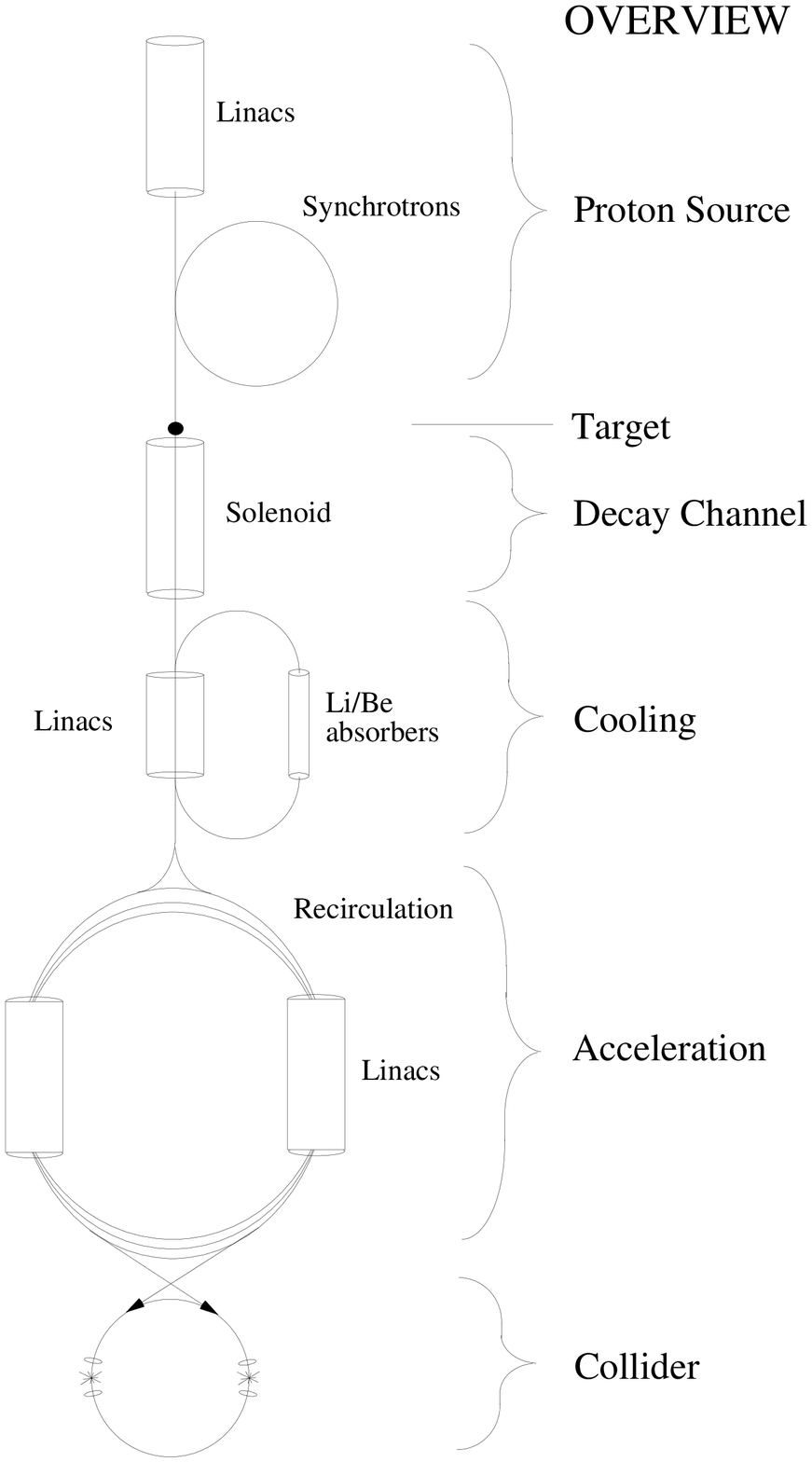,height=4.0in,width=3.5in}}
\label{schematic}
\end{figure}

The basic components of the \mumu collider are shown schematically in
Fig.\ref{schematic} and Tb.\ref{sum} shows parameters for the candidate designs.

A high intensity proton source  is bunch compressed and focussed on a heavy
metal target. The pions generated are captured by a high field solenoid and
transferred to a solenoidal decay channel within a low frequency linac. The
linac serves to reduce, by phase rotation the momentum spread of the pions, and of the muons into
which they decay. Subsequently, the muons are cooled by a
sequence of ionization cooling stages. Each stage consists of energy loss,
acceleration, and emittance exchange in energy absorbing wedges in the presence
of dispersion. Once they are cooled the muons must be rapidly accelerated to
avoid decay. This can be done in recirculating accelerators (\`{a} la CEBAF) 
or in
fast pulsed synchrotrons. Collisions occur in a separate high field collider
storage ring with very low beta insertion.
\begin{table} 
\begin{tabular}{llccc}
 & &4 TeV & 0.5 TeV & Demo \\
\tableline
Beam energy&TeV&2&0.25&0.25\\
Repetition rate & Hz & 15 & 15 & 2.5 \\
Muons per bunch & $10^{12}$ & 2 & 4 & 4 \\
Bunches of each sign &   & 2 & 1 & 1 \\
Norm. {\it rms} emittance $\epsilon^N$   & mm mrad  &  50  &  90 & 90\\
$\beta^*$ at intersection   & mm & 3 & 8 & 8 \\
Luminosity [units $10^{35}$] &${\rm cm}^{-2}{\rm s}^{-1}$&1 &$0.05$& $0.006$\\
\end{tabular}
\caption{Summary of Parameters of \mumu Colliders}
\label{sum}
\end{table}
\subsection{Physics Considerations}
There are at least two physics advantages of a \mumu collider, when compared
with a \ee:
   \begin{itemize}
\item Because of the lack of beamstrahlung, a \mumu collider can be
operated with an energy spread of as little as 0.01 \%. It is thus
possible to use the \mumu collider for precision measurements of masses and
widths, that would be hard, if not impossible, with an \ee collider.
\item The direct coupling of a lepton-lepton system to a Higgs boson has a
cross section that is proportional to the square of the mass of the lepton.
As a result, the cross section for direct Higgs production from the \mumu system
is
40,000 times that from an \ee system. \end{itemize}

However, there are liabilities:
\begin{itemize}
\item It is relatively hard to obtain both polarization and good luminosity in
a \mumu collider, whereas good polarization can be obtained in an \ee collider
without any loss in luminosity.
\item because of the decays of the muons, there will be a considerable
background of photons, muons and neutrons in the detector. This background may
be acceptable for some experiments, but it is certainly not as clean as an in
\ee collider.
   \end{itemize}
\subsection{Conclusion}
It is thus reasonable, from both technical and physics considerations, to
consider \mumu colliders as complementary to \ee colliders, just as \ee
colliders are complementary to hadron machines.
\section{SYSTEM COMPONENTS}
\subsection{Proton Driver}
The specifications of the proton drivers are given in Tb.\ref{driver}. In the
examples, the $\mu$-source driver is a high-intensity ($2.5\times  10^{13}$
protons per pulse) 30 GeV proton synchrotron. The preferred cycling rate would
be 15 Hz, but for the demonstration using the AGS\cite{roser}, the repetition
rate is limited to 2.5 Hz and to $24\,$GeV. For the lower energy machines, 2 final bunches are
employed (one to make $\mu^-$'s and the other to make $\mu^+$'s). For high
energy collider, four are used  (two $\mu$ bunches of each sign).

Earlier studies had suggested that the driver could be a 10 GeV machine with
the same charge per bunch, but a repetition rate of $30\,$Hz. This
specification was almost identical to that studied\cite{ref77} at ANL for a
spallation neutron source. Studies at FNAL\cite{mills} have further established
that such
a specification is not unreasonable. But in order to reduce the cost of the
muon phase rotation section and for minimizing the final muon longitudinal phase space, it
appears now that the final proton bunch length should be 1 ns (or even
less). This appears difficult to achieve at 10 GeV, but possible at 30 GeV.
If it is possible to obtain such short bunches with a 10 GeV
source, or if future optimizations allow the use of the longer bunches,
then the use of a lower energy source could be preferred.

In order to achieve the required 1 ns (rms) bunch length, an RF sequence
must be designed to phase rotate the bunch prior to extraction. The total
final momentum spread, based on the ANL parameters (95\% phase space of
$4.5\,{\rm eV s}$  per bunch), is $6\,\%,$  ($2.5\, \%$ rms), and the
space charge tune shift just  before extraction would be approx 0.5.
It might be necessary to perform this bunch compression in a separate
high field ring to avoid space charge problems.

  \begin{table} 
 \begin{tabular}{llccc}
                           &          & 4 TeV    & .5 TeV   &  Demo \\
\tableline
Proton energy              & GeV      &    30    &   30     &   24  \\
Repetition rate            & Hz       &    15    &    15    &    2.5 \\
Protons per bunch          & $10^{13}$&   2.5    &    2.5   &    2.5   \\
Bunches                    &          &   4      &    2     &    2   \\
Long. phase space/bunch    & eV s    &   4.5    &    4.5   &    4.5  \\
Final {\it rms} bunch length     & ns     &    1     &    1     &    1    \\
\end{tabular}
 \caption{Proton Driver Specifications}
\label{driver}
\end{table}
\subsection{Target and Pion Capture}
\begin{figure}[t!] 
\caption{ARC forward $\pi^+$ production vs proton energy and target material.}
\centerline{\epsfig{file=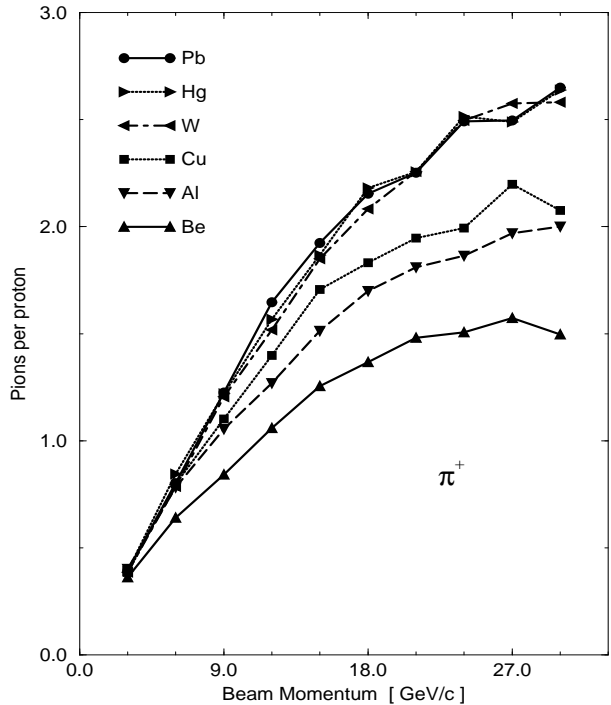,height=4.0in,width=3.5in}}
\vspace{10pt}
\label{pionproda}
\end{figure}

\begin{figure}[t!] 
\caption{ARC energy distribution for $24\,$GeV protons on Hg. }
\centerline{\epsfig{file=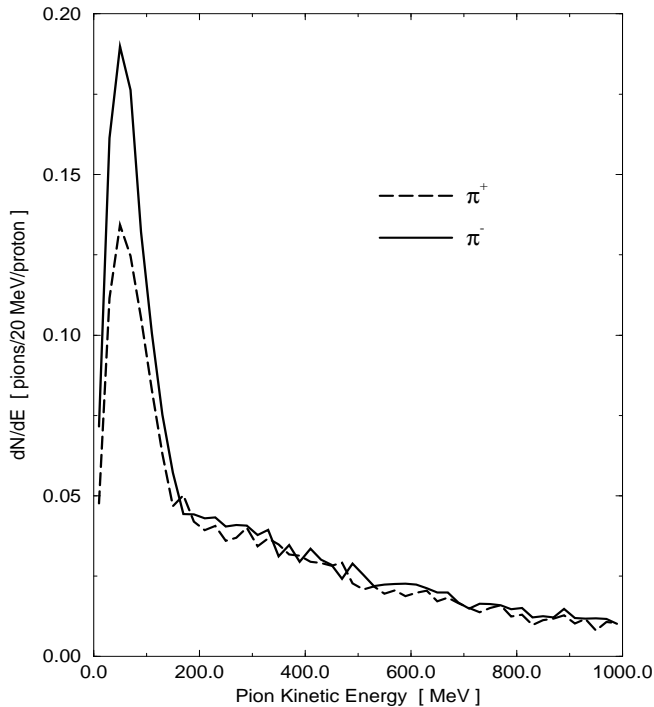,height=4.0in,width=3.5in}}
 \label{pionprodb}
 \end{figure}

Predictions of the nuclear Monte-Carlo program ARC\cite{arc} suggest that $\pi$
production is maximized by the use of heavy target materials, and that the
production is peaked at a relatively low pion energy ($\approx 100\,$MeV),
substantially independent of the initial proton energy. Fig. \ref{pionproda} 
shows the forward $\pi^+$ production as a function of proton energy and target
material; the $\pi^-$ distributions are similar. 
Fig.\ref{pionprodb} shows the energy
distribution for $\pi^+$ and $\pi^-$  for 24 GeV protons on Hg.

The target could be either Cu (approximately 24 cm long by 2 cm diameter), or
Hg (approximately 14 cm long by 2 cm diameter). A Hg target is being studied
for the European Spallation Source and would be cooled by circulating the
liquid. The Cu target would require water cooling.

Pions are captured from the target by a high-field ($20\,$T) water cooled
Bitter solenoid that surrounds it. Such a magnet is estimated\cite{weggel} to
require about 14 MW: a significant but not unreasonable power. The $\pi$'s are
then matched, using a suitable tapered field\cite{ref9}, into a periodic
superconducting solenoidal decay channel ($5\,$T and radius $=15\,$cm).

Monte Carlo studies indicate a yield of 0.6 muons, of each sign, per initial
proton, captured in the decay channel. But these pions have an extremely broad
energy spectrum so that only about half of them (0.3 $\mu/p$) can be used.
\subsection{Capture and Use of Both Signs}
Protons on the target produce pions of both signs, and a solenoid will capture
both, but the required subsequent RF systems will have opposite effects on
each. One solution is to break the proton bunch into two, target them on the
same target one after the other, and adjust the RF phases such as to act
correctly on one sign of the first bunch and on the other sign of the second.
This is the solution assumed in the parameters of this paper.

A second solution is to separate the pions of each charge prior to the use
of RF, and feed the beams of each charge into different channels. A third
possibility would be to separate the charges, delay the particles of one
charge, recombine the charges, and feed them into a single channel with
appropriate phases of RF.

After the target, and prior to the use of any RF or cooling, the beams have
very large emittances and energy spread. Conventional charge separation using
a dipole is not practical. But if a solenoidal channel is bent, then the
particles trapped within that channel will drift \cite{drift} in a direction
perpendicular to the bend. With our parameters this drift is dominated by a
term (curvature drift) that is linear with the forward momentum of the
particles, and has a  direction that depends on the sign of the charges.
It has been suggested \cite{snake} that if sufficient bend is employed, the
two charges could then
be separated by a septum and captured into two separate
channels. When these separate channels are bent back to the same forward
direction, then the momentum dispersion
is separately removed in each new channel.

Although this idea is very attractive, it has some problems:

\begin{itemize}
\item If the initial beam has a radius r=$0.15\,$m, and if the momentum range
to be accepted is $F={p_{{\rm max}}\over p_{{\rm min}}}=3,$ then the required
height of the solenoid just prior to separation is 2(1+F)r=$1.2\,$m. Use of a
lesser height will result in particle loss. Typically, the reduction in yield
for a curved solenoid compared to a straight solenoid is about $25\,\%$ (due to
the loss of very low and very high momentum pions), but this must be weighed
against the fact that both charge signs are captured for each proton on target.
\item The system of bend, separate, and return bend will require significant
length and must occur prior to the start of phase rotation (see below).
Unfortunately, it appears that the cost of the phase rotation RF appears to be
strongly dependent on keeping this distance as short as possible. On the other
hand it may be advisable to separate the remnant proton beam and other charged
debris exiting the target before the RF cavities. A curved solenoid would
accomplish this as well as charge-separate pions. 
 \end{itemize}
Clearly, compromises will be
involved, and more study of this concept is required.                          
\subsection{Phase Rotation Linac}
The pions, and the muons into which they decay, have an energy spread from
about 0 - 3 GeV, with an rms/mean of $\approx 100 \%$, and with a peak at
about 100 MeV. It would be difficult to handle such a wide spread in any
subsequent system. A linac is thus introduced along the decay channel, with
frequencies and phases chosen to deaccelerate the fast particles and
accelerate the slow ones; i.e. to phase rotate the muon bunch. Tb.\ref{rot}
gives an example of parameters of such a linac. It is seen that the lowest
frequency is 30 MHz: a low but not impossible frequency for a conventional
structure.
\begin{table}
\begin{tabular}{cccc}
Linac     & Length & Frequency & Gradient  \\
          &  m     &   MHz     &  MeV/m    \\
\tableline
1         &  3    &   60     &   5      \\
2         &  29   &   30     &   4       \\
3         &  5    &   60      &   4      \\
4         &  5    &   37     &   4       \\
\end{tabular}
\caption{Parameters of Phase Rotation Linacs}
\label{rot}
\end{table}

\begin{figure}[t!] 
\centerline{\epsfig{file=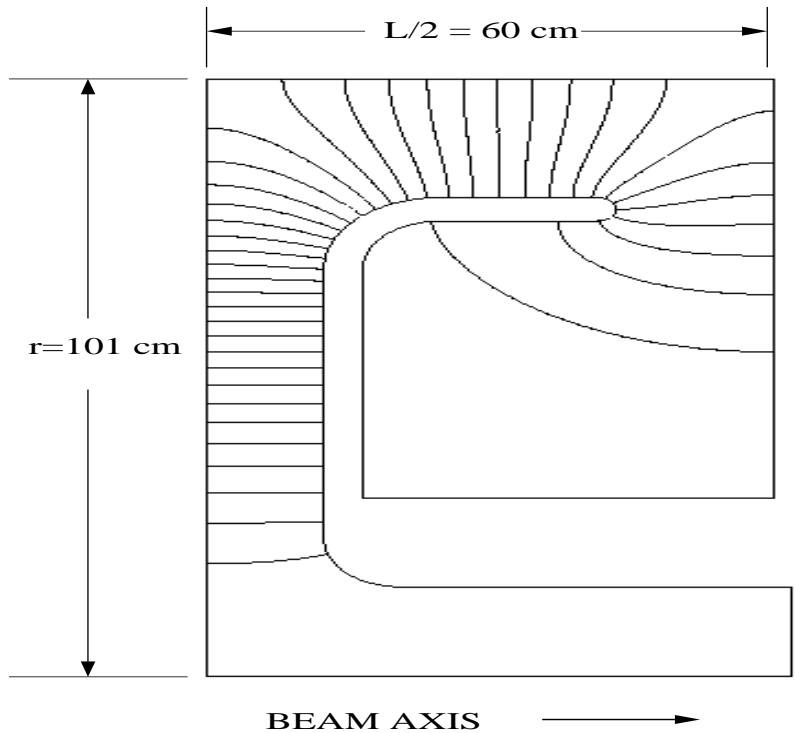,height=3.25in,width=3.25in}}
\vskip 1cm
\caption{30 MHz cavity for use in phase rotation and early stages of cooling.}
\label{30MHz}
\end{figure}

A design of a reentrant 30 MHz cavity is shown in Fig. \ref{30MHz}. Its
parameters are given in Tb. \ref{30MHzt}. It has a diameter of
approximately 2 m: only about one third that of a conventional pill-box 
cavity. To keep its cost down, it would be made of Al.
Multipactoring would probably be suppressed by stray fields from the 5 T
focusing coils, but could also be controlled by an internal coating of titanium
nitride.

\begin{table}
\begin{tabular}{llc}
Cavity Radius    & cm & 101 \\
Cavity Length  &   cm & 120  \\
Beam Pipe Radius  &  cm  &  15  \\
Accelerating Gap  &  cm  &  24  \\
Q               &        &  18200  \\
Average Acceleration Gradient & MV/m  &  3  \\
Peak RF Power     &  MW  &  6.3  \\
Average Power (15 Hz) & KW & 18.2  \\
Stored Energy   & J  &  609  \\
\end{tabular}
\caption{Parameters of 30 MHz RF Cavity}
\label{30MHzt}
\end{table}

After this phase rotation, a bunch can be selected with mean energy 150 MeV,
rms bunch length $1.7\,$m, and rms momentum spread  $20\,$\%. The number of
muons per initial proton in this selected bunch is 0.3. Unfortunately, the
linacs cannot phase rotate both signs in the same bunch: hence the need for
two bunches. The phases are set to rotate the $\mu^+$'s of one bunch and the
$\mu^-$'s of the other.

Prior to cooling, the bunch is accelerated to 300 MeV, at which energy the
momentum spread is 10 \%.
\subsection{Ionization Cooling}
\subsubsection{Cooling Theory}
For collider intensities, the phase-space volume must be reduced within the
$\mu$ lifetime. Cooling by synchrotron radiation, conventional stochastic
cooling and conventional electron cooling are all too slow. Optical stochastic
cooling\cite{ref11}, electron cooling in a plasma discharge\cite{ref12} and
cooling in a crystal lattice\cite{ref13} are being studied, but appear very
difficult. Ionization cooling\cite{ref14} of muons seems relatively
straightforward.

In ionization cooling, the beam loses both transverse and longitudinal momentum
as it passes through a material medium. Subsequently, the longitudinal
momentum can be restored by coherent reacceleration, leaving a net loss of
transverse momentum. Ionization cooling is not practical for protons and
electrons because of nuclear interactions (p's) and bremsstrahlung (e's) effects
in the material, but is practical for $\mu$'s because of their low nuclear
cross section and relatively low bremsstrahlung.

The equation for transverse cooling (with energies in GeV)  is:
  \begin{equation}
\frac{d\epsilon_n}{ds}\ =\ -\frac{dE_{\mu}}{ds}\ \frac{\epsilon_n}{E_{\mu}}\ +
\ \frac{\beta_{\perp} (0.014)^2}{2\ E_{\mu}m_{\mu}\ L_R},\label{eq1}
  \end{equation}
where $\epsilon_n$ is the normalized emittance, $\beta_{\perp}$ is the betatron
function at the absorber, $dE_{\mu}/ds$ is the energy loss, and $L_R$  is the
material radiation length.  The first term in this equation is the coherent
cooling term, and the second term is the heating due to multiple scattering.
This heating term is minimized if $\beta_{\perp}$ is small (strong-focusing)
and $L_R$ is large (a low-Z absorber).

From Eq.\ref{eq1} we find a limit to transverse cooling, which occurs when
heating due to multiple scattering  balances cooling due to energy loss. The
limits are $\epsilon_n\approx\ 0.6\ 10^{-2}\ \beta_{\perp}$ for Li, and
$\epsilon_n\approx\ 0.8\  10^{-2}\ \beta_{\perp}$ for Be.

The equation for energy spread  (longitudinal emittance) is:
 \begin{equation}
{\frac{d(\Delta E)^2}{ds}}\ =\
-2\ {\frac{d\left( {\frac{dE_\mu}{ds}} \right)} {dE_\mu}}
\ <(\Delta E_{\mu})^2 >\ +\
{\frac{d(\Delta E_{\mu})^2_{straggling}}{ds}}\label{eq2}
 \end{equation}
where the first term is the cooling (or heating) due to energy loss, 
and the second term is the heating due to straggling.

Cooling requires that  ${d(dE_{\mu}/ds)\over dE_{\mu}} > 0.$ But at energies
below about 200 MeV, the energy loss function for muons, $dE_{\mu}/ds$, is
decreasing with energy and there is thus heating of the beam.
Above 400 MeV the energy loss function increases gently, thus giving some
cooling, though not sufficient for our application.

In the long-path-length Gaussian-distribution limit, the heating term (energy
straggling)  is given by\cite{ref15}
 \begin{equation}
\frac{d(\Delta E_{\mu})^2_{straggling}}{ds}\ =\
4\pi\ (r_em_ec^2)^2\ N_o\ \frac{Z}{A}\ \rho\gamma^2\left(1-
\frac{\beta^2}{2}\right),
 \end{equation}
where $N_o$ is Avogadro's number and $\rho$ is the density. Since the energy
straggling increases as $\gamma^2$, and the cooling system size scales as
$\gamma$, cooling at low energies is desired.

Energy spread can also be reduced by artificially increasing
${d(dE_\mu/ds)\over dE_{\mu}}$ by placing a transverse variation in absorber
density or thickness at a location where position is energy dependent, i.e. where there is
dispersion. The use of such wedges can reduce energy spread, but it
simultaneously increases transverse emittance in the direction of the
dispersion. Six dimensional phase space is not reduced, but it does allow the
exchange of emittance between the longitudinal and transverse directions.
\subsubsection{Cooling System}
We require a reduction of the normalized transverse emittance by almost three
orders of magnitude (from $1\times 10^{-2}$ to $5\times 10^{-5}\,$m-rad), and a
reduction of the longitudinal emittance by more than one order of magnitude.
This cooling is obtained in a series of cooling stages. In general, each stage
consists of three components with matching sections between them:

 \begin{enumerate}
 \item a lattice consisting of spaced axial solenoids with alternating field
directions and lithium hydride
absorbers placed at the centers of the spaces between
them, where the $\beta_{\perp}$'s are minimum.
 \item a lattice consisting of more widely separated alternating solenoids,
and bending magnets to generate dispersion. At the location of maximum
dispersion, wedges of lithium hydride are introduced to interchange
longitudinal and transverse emittance;
 \item a short linac to restore the energy lost in the absorbers.
 \end{enumerate}

   At the end of this sequence of cooling stages, the transverse emittance can
be reduced to about $10^{-3}\,$m, still a factor of $\approx 20$ above the
emittance goals of Tb.\ref{sum}. The longitudinal emittance, however, can be
cooled to a value nearly three orders of magnitude less than  is required. The
additional reduction of transverse emittance can then be obtained by  a
reverse exchange of transverse and longitudinal phase-spaces. Again this is
done by the use of wedged absorbers in dispersive regions between solenoid
elements.

   Throughout this process appropriate momentum compaction and RF fields must
be used to control the bunch, in the presence of space charge, wake field and
resistive wall effects.

  In a few of the later stages, current carrying lithium rods might replace
item (1) above. In this case the rod serves simultaneously to maintain the low
$\beta_{\perp}$, and attenuate the beam momenta. Similar lithium rods, with
surface fields of $10\,$T , were developed at Novosibirsk and have been used
as focusing elements at FNAL and CERN \cite{ref16}). It is hoped\cite{20Tli}
that liquid lithium columns, can be used to raise the surface field to 20 T and improve
the resultant cooling.

It would be desirable, though not necessarily practicable, to economize on
linac sections by forming groups of stages into recirculating loops.
\subsubsection{Example}
A {\it model}
example has been generated that uses no lithium rods and no recirculating
loops. Individual components of the lattices used have been defined, but a
complete lattice has not yet been specified and no Monte Carlo study of its
performance has yet been performed. Spherical aberration due to solenoid end
effects, wake fields, and second order RF effects have not yet been included.

The phase advance in each cell of the lattice is made as close to $\pi/2$ as
possible in order to minimize the $\beta$'s at the location of the absorber,
but it is kept somewhat less than this value so that the phase advance per
cell should never exceed $\pi/2.$ The following effects are included:
 \begin{enumerate}
 \item the maximum space charge transverse defocusing
 \item a 3 $\sigma$ fluctuation of momentum
 \item a 3 $\sigma$ fluctuation in amplitude
\end{enumerate}

   In all but the final stages of cooling it is assumed that both charges will
use the same channel. Bending magnets are introduced to generate dispersion,
but the dispersion is kept equal to zero at the center of all solenoids. The
maximum allowed beam angle with respect to the axis, due to dispersion, is 45
degrees.

   In the early stages, the solenoids are relatively large and their fields
are limited to $7\,$T. In later stages the emittance has decreased, the
apertures are smaller and the fields are increased to $10\,$T. The maximum
bending fields used are $7\,$T, but most are at $3\,$T.

\begin{figure}[t!] 
\caption{$\epsilon_{\perp}$, $\epsilon_{L}$ and E$_{\mu}$ [GeV], vs stage number in the cooling sequence. }
\centerline{\epsfig{file=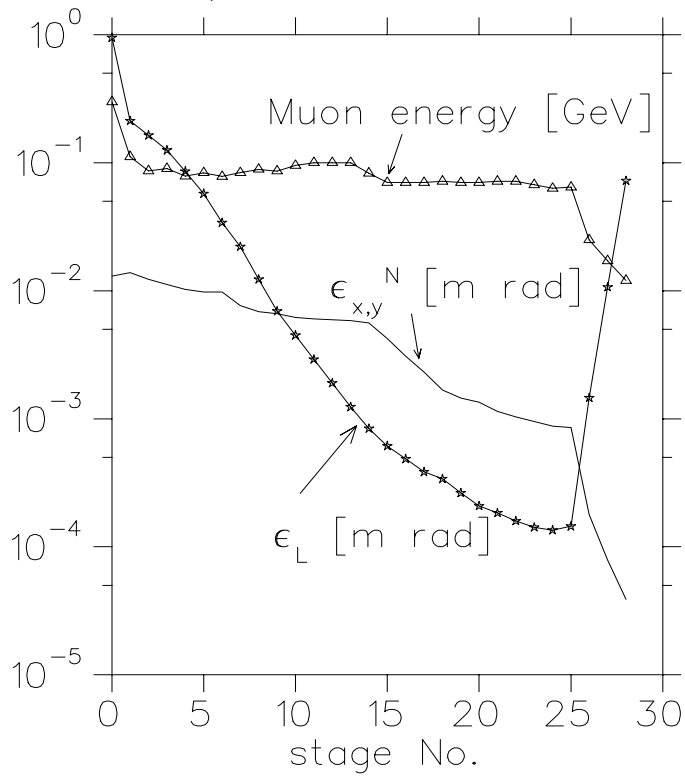,height=4.0in,width=3.5in}}
 \label{cooling}
 \end{figure}

The emittances, transverse and longitudinal, as a function of stage number, are
shown in Fig.\ref{cooling}, together with the beam energy. In the first 15
stages, relatively strong wedges are used to rapidly reduce the longitudinal
emittance, while the transverse emittance is reduced relatively slowly. The
object here is to reduce the bunch length, thus allowing the use of higher
frequency and higher gradient RF in the reacceleration linacs. In the next 10
stages, the emittances are reduced close to their asymptotic limits. The
charges are now separated for the last two stages. In these stages, the
transverse and longitudinal emittances are again exchanged, but in the opposite
direction: lowering the transverse and raising the longitudinal. During this
exchange the energy is allowed to fall to 10 MeV in order to minimize the
$\beta$, and thus limit the emittance dilution.

   The total length of the system is 880 m, and the total acceleration used is
2.7 GeV. The fraction of muons remaining at the end of the cooling system
is calculated to be $43\,$\%.

\subsection{Acceleration}
Following cooling and initial bunch compression the beams must be accelerated
to full energy (2 TeV, or 250 GeV). A sequence of linacs would work, but would
be expensive. Conventional synchrotrons cannot be used because the muons would
decay before reaching the required energy. The conservative solution is to use
a sequence of recirculating accelerators (similar to that used at CEBAF). A
more economical solution would be to use fast rise time pulsed magnets in
synchrotrons, or synchrotrons with rapidly rotating permanent magnets
interspersed with high field fixed magnets.
\subsubsection{Recirculating Acceleration}
\begin{figure}[t!] 
\caption{A cross section of a 9 aperture sc magnet.  }
\centerline{\epsfig{file=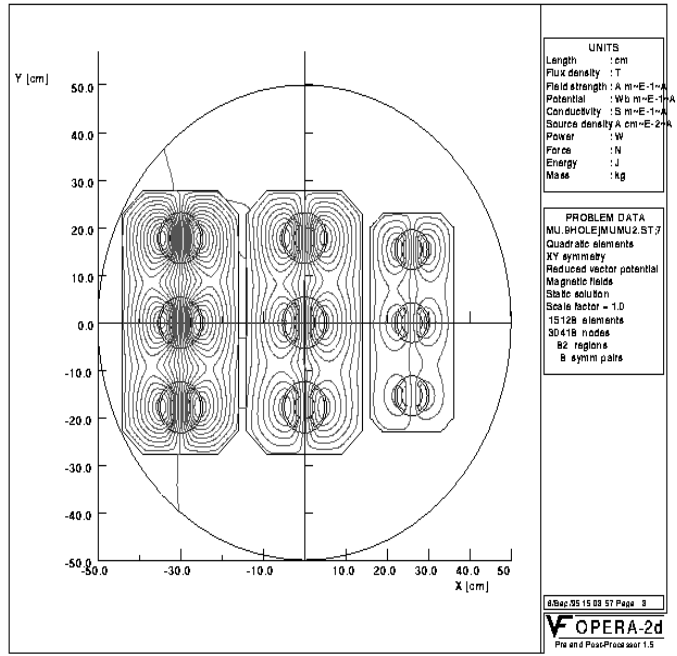,height=4.0in,width=4.5in}}
 \label{9hole}
 \end{figure}

Tb.\ref{acceleration} gives an example of a possible sequence of
recirculating accelerators. After an initial linac from 0.2 $\rightarrow$ 1
GeV, there are two conventional RF recirculating accelerators taking the muons
up to 75 GeV, then two superconducting recirculators going up to 2000 GeV.
\begin{table}
 \begin{tabular}{llccccc}
                     &         & Linac &  \#1   &  \#2   &   \#3  &   \#4  \\
\tableline
initial energy       & GeV     &   0.20&     1 &     8 &    75 &   250 \\
final energy         & GeV     &     1 &     8 &    75 &   250 &  2000 \\
nloop                &         &     1 &    12 &    18 &    18 &    18 \\
freq.                & MHz     &   100 &   100 &   400 &  1300 &  2000 \\
linac V              & GV      &   0.80&   0.58&   3.72&   9.72&  97.20 \\
grad                 &         &     5 &     5 &    10 &    15 &    20  \\
dp/p  initial        & \%      &    12 &   2.70&   1.50&     1 &     1  \\
dp/p  final          & \%      &   2.70&   1.50&     1 &     1 &   0.20 \\
$\sigma_z$ initial   & mm      &   341 &   333 &  82.52&  14.52&   4.79 \\
$\sigma_z$ final     & mm      &   303 &  75.02&  13.20&   4.36&   3.00 \\
$\eta$               & \%      &   1.04&   0.95&   1.74&   3.64&   4.01 \\
$N_\mu$              & $10^{12}$ &   2.59&   2.35&   2.17&   2.09&   2    \\
$\tau_{fill}$        & $\mu$s  &  87.17&  87.17&  10.90&  s.c. &  s.c.  \\
beam t               & $\mu$s  &   0.58&   6.55&  49.25&   103 &   805   \\
decay survival       &         &   0.94&   0.91&   0.92&   0.97&   0.95  \\
linac len            & km      &   0.16&   0.12&   0.37&   0.65&   4.86  \\
arc len              & km      &   0.01&   0.05&   0.45&   1.07&   8.55  \\
tot circ             & km      &   0.17&   0.16&   0.82&   1.72&  13.41  \\
phase slip           & deg     &     0 &  38.37&   7.69&   0.50&   0.51  \\
\end{tabular}
 \caption{Parameters of Recirculating Accelerators} 
\label{acceleration}
\end{table}
   Criteria that must be considered in picking the parameters of such
accelerators are:
\begin{itemize}
\item The wavelengths of rf should be chosen to limit the loading, $\eta$, (it
is restricted to below 4 \% in this example) to avoid excessive longitudinal
wakefields and the resultant emittance growth.
 \item  The wavelength should also be sufficiently large compared to the bunch
length to avoid second order effects (in this example: 10 times).
 \item  For power efficiency, the cavity fill time should be
long compared to the acceleration time. When conventional cavities cannot
satisfy this condition, superconducting cavities are required.
 \item  In order to minimize muon decay during acceleration (in this example
73\% of the muons are accelerated without decay), the number of
recirculations at each stage should be kept low, and the RF acceleration
voltage correspondingly high. But for the minimum cost the number of
recirculations appears to be of the order of 20 - a relatively high number. In
order to avoid a large number of separate magnets, multiple aperture magnets
can be designed (see Fig.\ref{9hole}).
 \end{itemize}

Note that the linacs see two bunches of opposite signs, passing through
in opposite directions. In the final accelerator in the 2 TeV case, each bunch
passes through the linac 18 times. The total loading
is then $4\times 18\times \eta = 288 \%.$ With this loading, assuming 60\%
klystron efficiencies and reasonable cryogenic loads, one could probably
achieve 35\% wall to beam power efficiency, giving a wall power
consumption for the RF in this ring of 108 MW.

A recent study\cite{neufferacc} tracked particles through a
similar sequence of recirculating accelerators and found a dilution of
longitudinal phase space of the order of 15\%.
\subsubsection{Pulsed Magnet Synchrotron Alternatives}
\begin{figure}[t!] 
\caption{Cross section of pulsed magnet for use in the
acceleration to 250 GeV.  }
\centerline{\epsfig{file=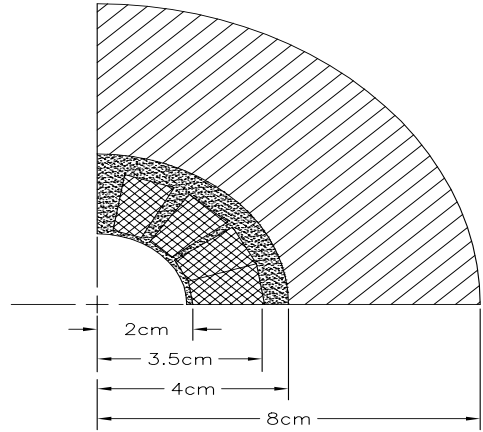,height=4.0in,width=3.5in}}
 \label{pulse}
 \end{figure}

An alternative to recirculating accelerators for stages \#2 and \#3 would be to
use pulsed magnet synchrotrons. The cross section of a pulsed magnet for this
purpose is shown in Fig. \ref{pulse}. If desired, the number of recirculations
could be higher in this case, and the needed RF voltage correspondingly lower,
but the loss of particles from decay would be somewhat more. The cost for a
pulsed magnet system appears to be significantly less than that of a multi-hole
recirculating magnet system, but its power consumption seems impractical at
energies above 250 GeV.
\subsubsection{Pulsed and Rotating Magnet Alternatives for the Final
Accelerator}
For the final acceleration to 2 TeV in the high energy machine, the power
consumed by pulsed magnets would be excessive. A recirculating accelerator is
still usable, but there are two other, possibly cheaper, alternatives being
considered:

   a) A sequence of two hybrid accelerators (0.25-1, and 1-2 TeV), each
employing superconducting fixed magnets (e.g. $10\,$T) alternating with pairs of
counter-rotating permanent magnets\cite{rotating}. The power consumption
would be negligible, but its practicality is not yet clear.

   b) A similar sequence of two hybrid accelerators (0.25-1, and 1-2 TeV),
each again employing alternate superconducting fixed magnets (e.g. $10\,$T), but
instead of pairs of rotating magnets, pulsed warm magnets (whose fields might
swing from -1.5 T to +1.5 T) would be used. The power consumption would be
considerable, but the initial cost might be significantly less than that for
a recirculating accelerator, and it might be more practical than the rotating
magnet scheme.
\subsection{$\mu$ Storage Ring}
After acceleration, the $\mu^+$ and $\mu^-$ bunches are injected into a
separate storage ring. The highest possible average bending field is
desirable, to maximize the number of revolutions before decay, and thus
maximize the luminosity. Collisions would occur in one, or perhaps two, very
low-$\beta^*$ interaction areas. Parameters of the rings are given in
Tb.\ref{collider}.
  \begin{table}  
 \begin{tabular}{llccc}
                           &          & 4 TeV & .5 TeV & Demo.  \\
\tableline
Beam energy                & TeV      &     2    &   .25  &  .25  \\
Beam $\gamma$              &          &   19,000 &  2,400   &  2,400 \\
Repetition rate            & Hz       &    15    &    15    &    2.5 \\
Muons per bunch            & $10^{12}$  &   2    &    4     &    4   \\
Bunches of each sign       &          &   2      &    1     &    1   \\
Normalized {\it rms} emittance $\epsilon^N$   &mm mrad  &  50  &  90   &   90   \\
Bending Field              &  T   &    9    &    9    &    8   \\
Circumference              &  Km      &    7    &    1.2  &   1.5   \\
Average ring mag. field $B$    & T   & 6    &   5     &    4  \\
Effective turns before decay &       & 900    &   800  &   750 \\
$\beta^*$ at intersection   & mm     &   3   &   8   &   8    \\
{\it rms} beam size at I.P.       & $\mu m$&   2.8 &  16   &   16   \\
\smallskip
Luminosity &${\rm cm}^{-2}{\rm s}^{-1}$& $10^{35}$&5\ $10^{33}$& $6\ 10^{32}$\\
\end{tabular}
 \caption{Parameters of Collider Rings}
\label{collider}
\end{table}

  The bunch populations decay exponentially, yielding an
integrated luminosity equal to its initial value multiplied by an
{\it effective} number of turns $n_{effective}=150\ B,$ where B is the mean
bending field in T.
\subsubsection{Bending Magnet Design}
\begin{figure}[b!] 
\caption{Energy attenuation vs the thickness of a tungsten liner. }
\centerline{\epsfig{file=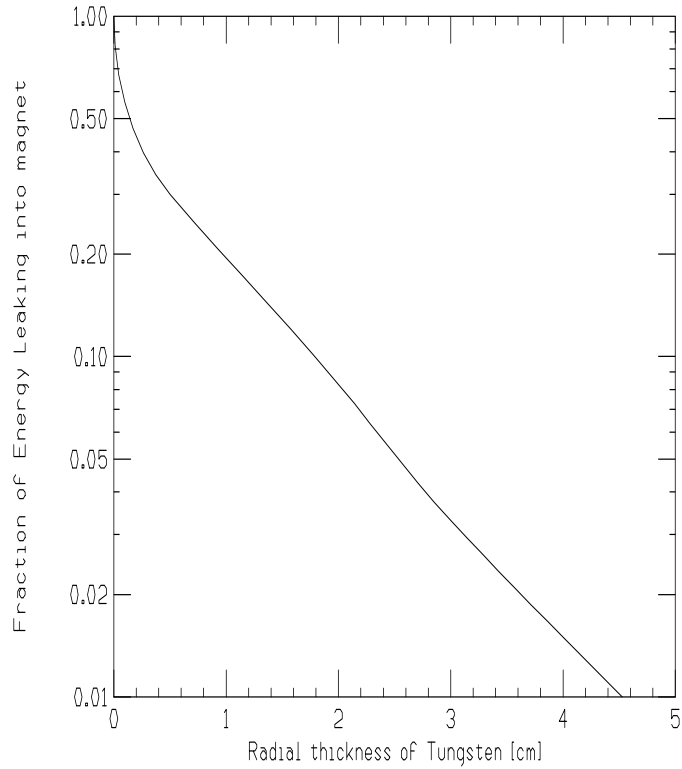,height=4.0in,width=3.5in}}
 \label{shield}
 \end{figure}

   The magnet design is complicated by the fact that the $\mu$'s decay within
the rings ($\mu^-\ \rightarrow\ e^-\overline{\nu_e}\nu_{\mu}$), producing
electrons whose mean energy is approximately 1/3 that of the muons. These
electrons travel to the inside of the ring dipoles, radiating a substantial
fraction of their energy as synchrotron radiation towards the outside of the
ring. Fig.\ref{shield} shows the attenuation of the heating produced as a
function of the thickness of a warm tungsten liner.
If conventional superconductor is used, then the thicknesses required in the
three cases would be as given in Tb.\ref{linert}. If high Tc
superconductors could be used then these thicknesses could probably be halved.

 \begin{table}  
 \begin{tabular}{llccc}
                  &        &  2TeV   &   0.5 TeV   &    Demo   \\
\tableline
Unshielded Power  &   MW   &  17     &    4        &    .5   \\
Liner thickness  &   cm   &   4.5   &    3        &     2    \\
Power leakage     &   KW   &   170   &    150      &     50   \\
\end{tabular}
 \caption{Required Thickness of Shielding in Collider Magnets.}
\label{linert}
\end{table}
\subsubsection{Lattice Design}
  Studies\cite{stability} of the  resistive wall impedance instabilities
indicate that the required muon bunches
(eg for $2\,$TeV: $\sigma_z=3\ mm,\ N_{\mu}= 2\times 10^{12}$) would be unstable
in a conventional ring. In any case, the RF requirements to maintain such
bunches would be excessive. It is thus proposed to use an isochronous lattice of
the type discussed by S.Y. Lee {\it et al}\cite{ref17}. The elements of such a
lattice have been designed\cite{Djan}, and are being incorporated into a full
ring.

It had been feared that amplitude dependent anisochronicity generated in the
insertion would cause bunch growth in an otherwise purely isochronous design.
It has, however, been pointed out \cite{oide}
that if chromaticity is corrected in the
ring, then amplitude dependent anisochronicity is automatically removed.
\subsubsection{Low $\beta$ Insertion}
\begin{figure}[t!] 
\caption{{\it rms} radius of the beam at the last four quadrupoles of the 
final focus. }
\centerline{\epsfig{file=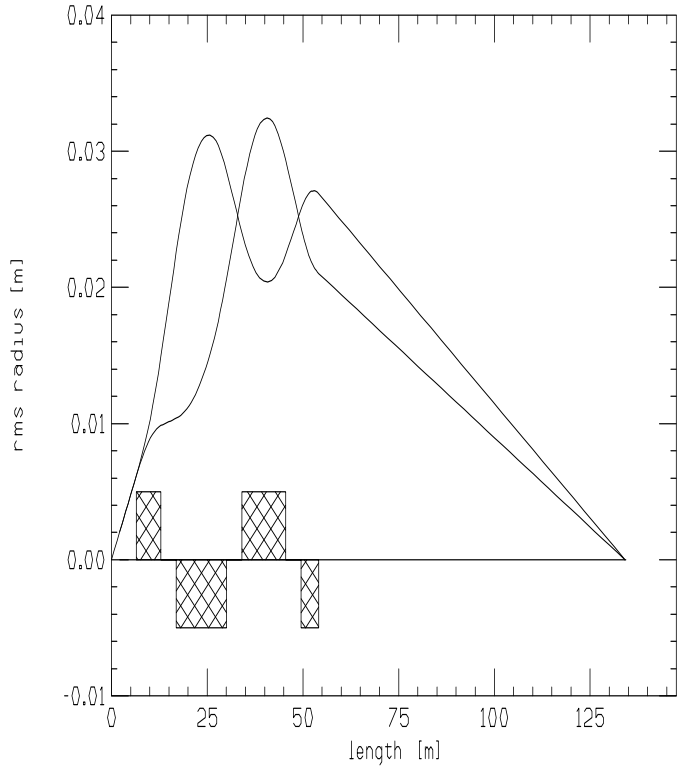,height=4.0in,width=3.5in}}
 \label{ff}
 \end{figure}

In order to obtain the desired luminosity we require a very low beta at the
intersection point: $\beta^*=3\,{\rm mm}$ for 4 TeV, $\beta^*=8\,{\rm mm}$ for .5 TeV. A
possible final focusing quadruplet design is shown in Fig.\ref{ff}. The
parameters of the quadrupoles for this quadruplet are given in
Tb.\ref{ffquads}. With these elements, the maximum beta's in both x and y are
of the order of 400 km in the 4 TeV case, and 14 km in the 0.5 TeV machine.
The
chromaticities $(1/4\pi\int \beta dk)$ are approximately 6000 for the 4 TeV
case, and 600 for the .5 TeV machine. Such chromaticities are too large to
correct within the rest of a conventional ring and therefore require
local correction of\cite{ref18}.
\begin{table}
 \begin{tabular}{lccccc}
  & &\multicolumn{2}{c}{$4\,$MeV}& \multicolumn{2}{c}{$0.5\,$MeV} \\
\tableline
       & field (T) &  L(m) & R(cm)& L(m) & R(cm)  \\
\tableline
drift  &      & 6.5    &    &   1.99   &         \\
focus  &   6  &  6.43  & 6  &   1.969  & 5.625   \\
drift &       & 4.0    &    &   1.2247 &         \\
defocus & 6.4 & 13.144 & 12 &   4.025  & 11.25  \\
drift &       &  4.0    &    &  1.2247 &         \\
focus  &  6.4 & 11.458 & 12 &   3.508  & 11.25  \\
drift &       &   4.0   &    &  1.2247 &         \\
defocus &6.348& 4.575  & 10 &   1.400  & 9.375  \\
drift   &     &  80    &     & 24.48        &             \\
\end{tabular}
 \caption{Final Focus Quadrupoles; L and R are the length and the radius
respectively. } 
\label{ffquads}
\end{table}

\begin{figure}[t!] 
\caption{Dispersion and $\beta$ functions for different momenta ($\delta p/p=0.4\,\%$).  }
\centerline{\epsfig{file=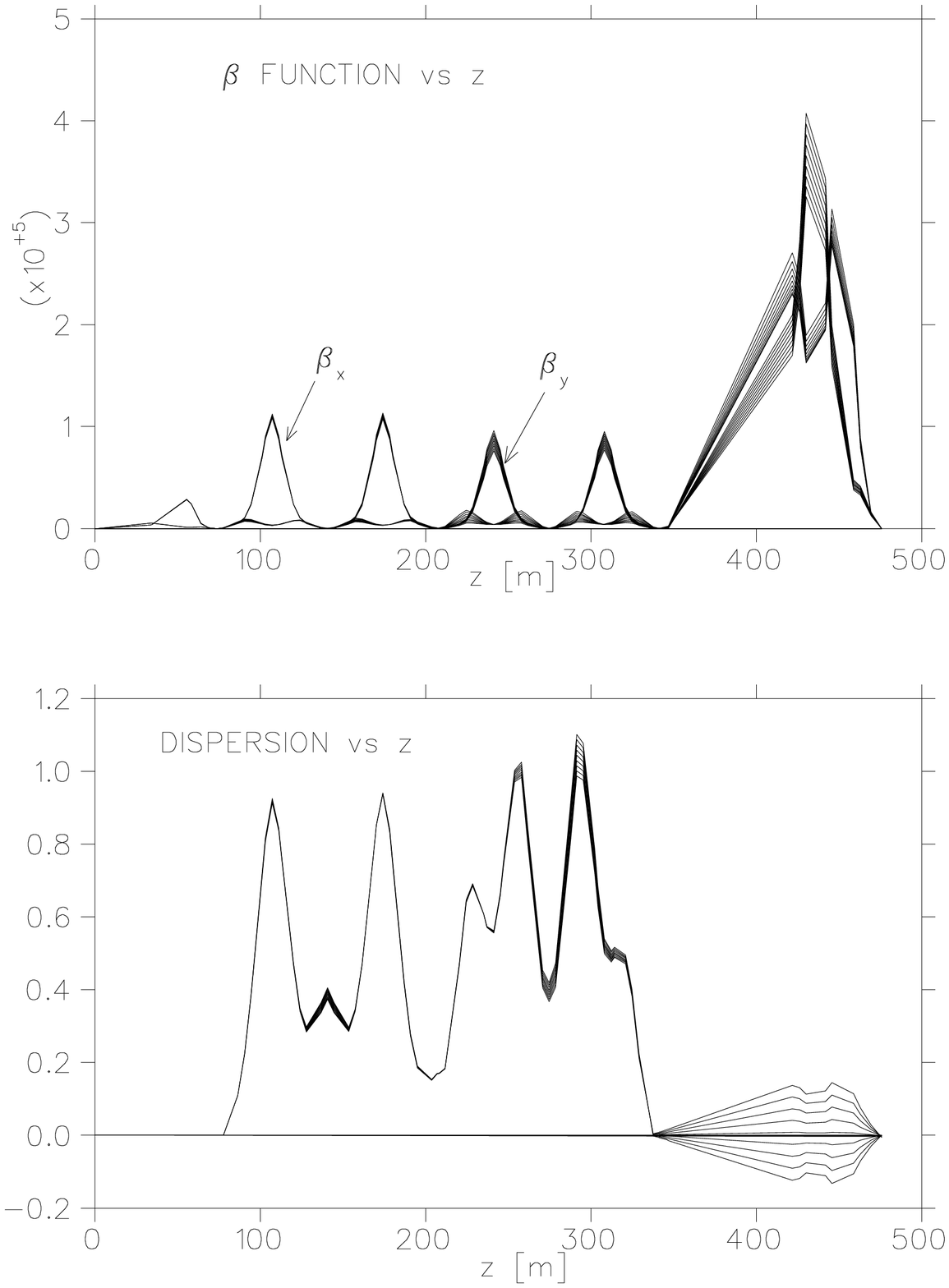,height=4.0in,width=3.5in}}
 \label{irbetas}
 \end{figure}

\begin{figure}[t!] 
\caption{The tune shift as a function of momentum for the
{\em model} insertion design.  }
\centerline{\epsfig{file=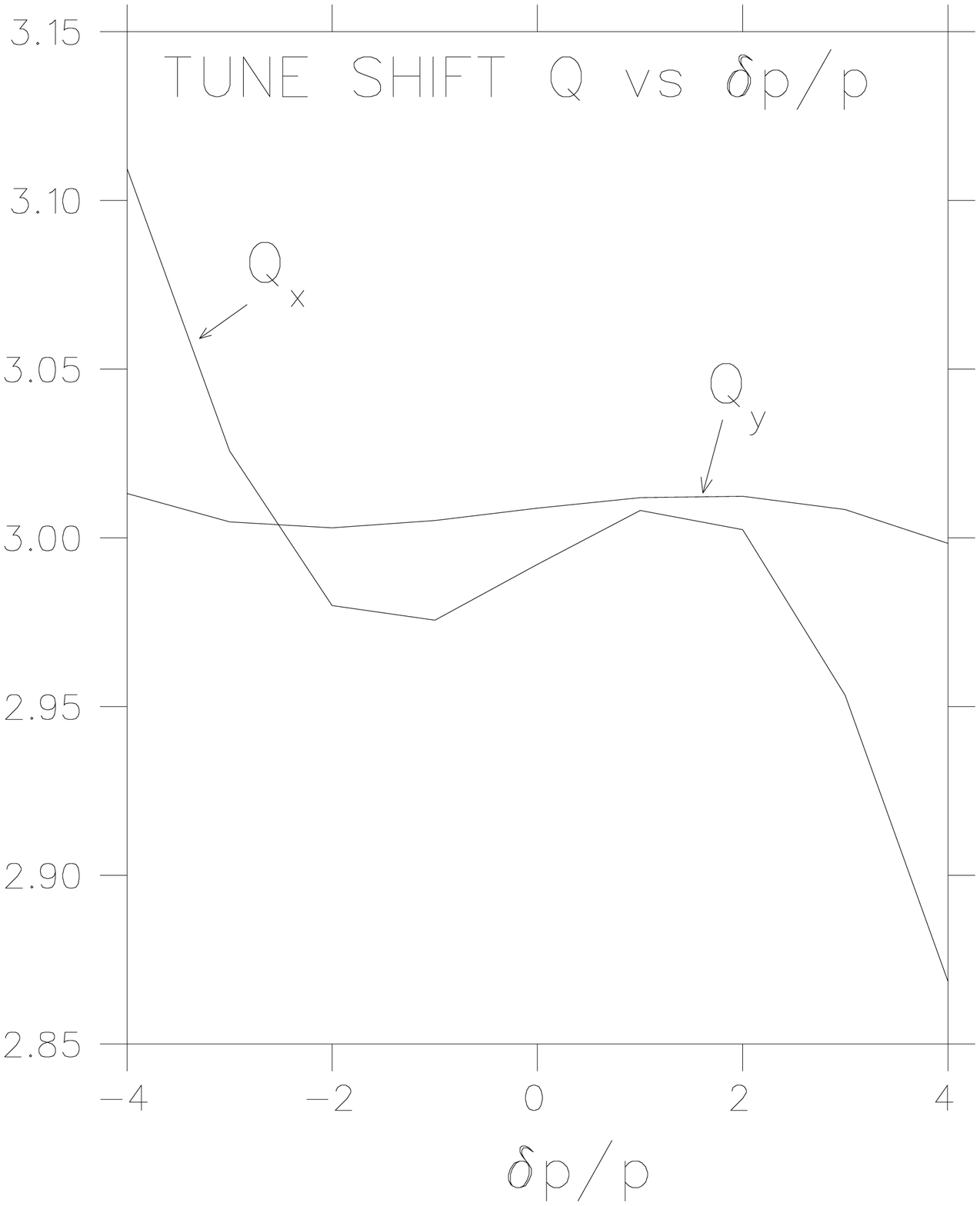,height=4.0in,width=3.5in}}
 \label{irtune}
 \end{figure}

   A preliminary {\it model} design\cite{ff} of local chromatic correction (for
the 4 TeV case) has been presented. Fig.\ref{irbetas} shows the horizontal
dispersion and beta functions for this design.
 Fig.\ref{irtune} shows the
tune shift as a function of momentum. It is seen that this design has a
momentum acceptance of $\pm 0.3\,\%$. The second order amplitude dependent
tune shifts appear acceptable in this design, but the bending fields are
unrealistic.
It is expected that these limitations will soon be overcome, and that more
sophisticated designs\cite{ref19} should do even better. It is hoped to
achieve a momentum acceptance of $\pm 0.6\,\%$ for use with a clipped rms
momentum spread of 0.2 \%.

\section{COLLIDER PERFORMANCE}
\subsection{Luminosity, Energy and Momentum Spread}
The luminosity is given by:
 \b
\Ls\={N^2\ f\ n_e \gamma\over 4\pi\ \beta^*\ \epsilon_n} H(A,D)
 \e
where $A=\sigma_z / \beta^*,$ $D={2\sigma_z N\over \gamma
\sigma_{x,y}(\sigma_x+\sigma_y)}r_e ({m_e\over m_{\mu}})$ and the enhancement
factor is $
H(A,D)\approx 1+D^{1/4} \left[ {D^3\over 1+D^3} \right] \left\{
\ln{(\sqrt{D}+1)} + 2\ln{({0.8\over A})} \right\}.$

    The luminosities given in Tb.\ref{sum} are those for the design energy
and energy spread. At lower energies, or energy spread, the luminosities will
be lower.

For a fixed collider lattice, operating at energies lower than the
design value, the luminosity will fall as $\gamma^3.$ One
power comes from the $\gamma$ in the above equation; a second comes from
$n_e$, the effective number of turns, that is proportional to
${\gamma\over 2\pi R}$; the third term comes from $\beta^*,$ which must be
increased proportional to $\gamma$ in order to keep the beam size constant
within the focusing magnets. The bunch length $\sigma_z$ must also be increased
so that the required longitudinal phase space is not decreased.

In view of this rapid drop in luminosity with energy, it would be desirable
to have separate collider rings at relatively close energy spacings: e.g.
not more than factors of two apart.

   If it is required to lower the energy spread $\Delta E/E$ at a fixed
energy, then, given the same longitudinal phase space, the bunch length
$\sigma_z$ must be increased. If the final focus is retuned to simultaneously
increase $\beta^*$ to maintain the value of $A,$ then the luminosity will be
exactly proportional to  $\Delta E/E.$ But if, instead, the $\beta^*$ is kept
constant, and the parameter A allowed to increase, then the luminosity falls
initially at a somewhat lower rate. The luminosity, for small  $\Delta E/E$ is
then approximately given by:
 \b
\Ls\=2\ \Ls_0\ {\Delta E/E\over {\Delta E/E}_0}.
 \e
There may, however, be tune shift emittance growth problems in this case.
\subsection{Detector Background}
There will be backgrounds from the decay of muons in the ring, from muon halo
around the beam, and there will be backgrounds from the interactions themselves.
\subsubsection{Muon Decay Background}
  A first Monte Carlo study\cite{ref20} of the muon decay background was done with the MARS95
code\cite{ref21}, based on a preliminary insertion lattice. A tungsten
shielding {\it nose} was introduced, extending to within 15 cm of the
intersection point. It was found that:
\begin{itemize}
   \item{a large part of the electromagnetic background came from synchrotron
radiation, due to the bending magnets in the chromatic correction
section.
   \item As many as 500 hits per cm$^2$ were expected in a vertex detector,
falling to of the order of 2 hits per cm$^2$ in an outer tracker. }
   \item{There was considerable, very low energy neutron background: of the
order of 30,000 neutrons per $cm^2,$ giving, with an efficiency of $3\ 10^{-
3}$, about 100 hits per cm$^2$. }
 \end{itemize}

   It was hoped that by improving the shielding these backgrounds could be
substantially reduced.

   A more recent study\cite{iuliu} of the electromagnetic component of the
background has been done using the GEANT codes\cite{geant}. This study
differed from the first in several ways:
\begin{itemize}
\item{ the shower electrons and photons were followed down to a
lower energy ($50\,$keV for electrons and $15\,$keV for photons).}

\item{The {\it nose} angle, i.e. the angle not seen by the detector, was
increased
from 9 to 20 degrees.}

\item{The {\it nose} design was modified (see Fig.\ref{nose}) so that: 1) The
incoming electrons are collimated to $\pm\ 4\ \times\ \sigma_{\theta_0}$ (where
$\sigma_{\theta_0}$ is the rms divergence of the beam) by a cone leading down towards the
vertex. 2)  The detector could not {\it see} any
surface
directly illuminated by these initial electrons, whether seen in the forward
or backward (albedo) directions. 3)  The detector could not {\it see } any
surface
that is illuminated by secondary electrons if the secondary scattering angle is forward.
4)  The minimum distance through the collimator from the detector to any
primarily illuminated surface was more than 100 mm, and from any
secondarily illuminated surface, 30 mm.                      }
\begin{figure}[t!] 
\centerline{\epsfig{file=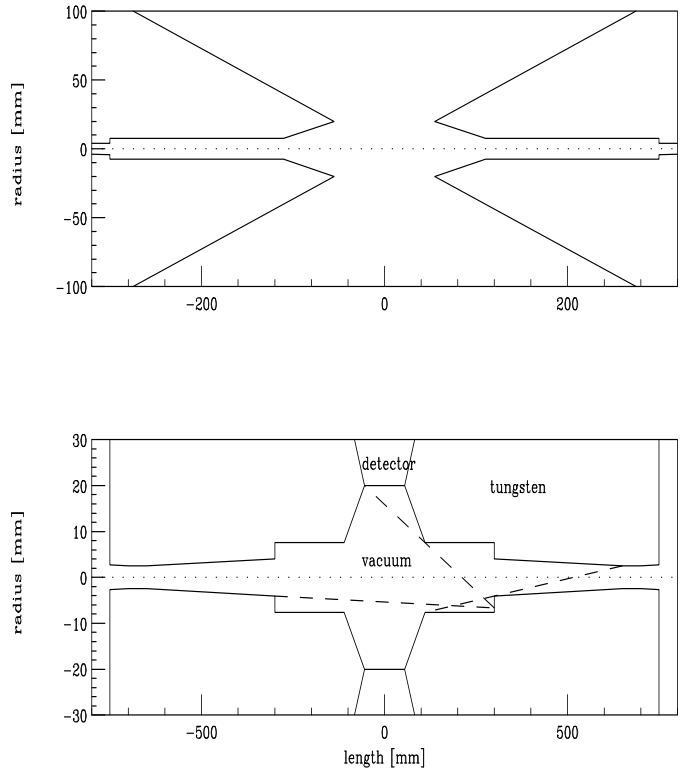,height=4.0in,width=3.5in}}
\caption{Schematic of the detector nose.}
 \label{nose}
 \end{figure}

\item{It was assumed that a collimator placed at a focus  130 m from the intersection point would
be able to effectively shield all synchrotron photons from the bending magnets
beyond that point (the {\it rms} beam
size at this focus is only 10 $\mu m$). }
 \end{itemize}

This study indicated that the dominant background was not from synchrotron
photons, but from $\mu$ decay electrons. The average momentum of these photons
was only 1 MeV. and the sensitivity of detectors to such low energy photons is
only about 0.3\% for silicon detectors and as low as 0.1\% for a suitably
designed gas detector. Tb.\ref{background} gives the total numbers of photons,
the total number of hits, possible pixel sizes, and the hits per pixel, for a)
a vertex detector placed at a 5 cm radius, and b) a gas detector placed at a 1
m radius. In all cases the numbers are given per bunch crossing.
\begin{table}
\begin{tabular}{lcc}
Detector              &   vertex     &   tracker  \\
Radius                &      5 cm      &    1 m   \\
\tableline
Number of photons     & $50\ 10^6$   & $15\ 10^6$    \\
Number of hits        &   150,000     &  15,000      \\
Detector Area         &   863 cm$^2$  &  34 m$^2$   \\
Pixel size            & 20 x 20 $\mu$m& 1 mm x 1 cm    \\
Sensitivity           &   0.3 \%     &    0.1 \%      \\
Occupancy             & .07 \%       &    0.4 \%     \\
\end{tabular}
\caption{Detector Backgrounds from $\mu$ decay}
\label{background}
\end{table}
This study found a relatively modest flux of muons from $\mu$ pair production
in electromagnetic showers: about 50 such tracks pass through the detector per
bunch crossing.

   The general conclusion of this study is not inconsistent with that from the
MARS study: the background, though serious, is not apparently impossible.
Further reductions are expected as the shielding is optimized, and it may be
possible to design detectors that are less sensitive to the very low energy
neutrons and photons present.
\subsubsection{Muon Halo Background}
   There would be a very serious background from the presence of even a very
small halo of near full energy muons in the circulating beam. The beam will
need careful preparation before injection into the collider, and a collimation
system will be designed to be located in the ring, possibly on the opposite
side from the detector.
\subsubsection{Electron Pair Background}
   In \ee machines there is a significant problem from beamstrahlung photons
(synchrotron radiation of beam particles in the coherent field of the oncoming
bunch), and an additional problem from pair production from these photons.

   With muons, there is negligible beamstrahlung, and thus negligible pair
production from these real photons. Pisin Chen\cite{beamstrahlung} has further shown
that beamstrahlung of electrons from the nearby decay of muons does not pose a
problem.

There is, however, significant incoherent (i.e. \mumu $\rightarrow$ \ee) pair
production in the 4 TeV Collider case. The cross section is estimated to be
$10\,$mb\cite{pairsection}, which would give rise to a background of $\approx
3\,10^4$  electron pairs per bunch crossing. Most of these, approximately 
$90\,\%$, will be trapped
inside the tungsten nose cone, but those with
energy between 30 and $100\,$MeV will enter the detector region.

   There remains some question about the coherent pair production generated
by the virtual photons interacting with  the coherent electromagnetic fields of the entire oncoming bunch. A simple
Weizs\"acker-Williams calculation\cite{pisin} yields a background so disastrous
that it would consume the entire beam at a rate comparable with its decay.
However, I. Ginzburg\cite{ginzburg} and others have argued that the integration
must be cut off due to the finite size of the final electrons. If this is true,
then the background becomes negligible.

   If the coherent pair production problem is confirmed, then there are two possible
solutions:
   1) one could design a two ring, four beam machine (a $\mu^+$ and a
$\mu^-$ bunch coming from each side of the collision region, at the same
time). In this case the coherent electromagnetic fields at the intersection
are canceled and the pair production becomes negligible.
   2) Plasma could be introduced at the intersection point to cancel the beam
electromagnetic fields\cite{plasma}.
\section{CONCLUSION}
 \begin{itemize}
 \item The scenario for a 2 + 2 TeV, high luminosity collider is by no means
complete. There are many problems still to be examined. Much work remains to
be done, but no obvious show stopper has yet been found.
 \item Many technical components require development: a large high field
solenoid for capture, low frequency RF linacs, long lithium lenses, multi-beam
pulsed and/or rotating magnets for acceleration, warm bore shielding inside
high field dipoles for the collider, muon collimators and background shields,
but:
 \item None of the required components may be described as {\it exotic}, 
and their specifications are not far beyond what has been demonstrated.
 \item If the components can be developed and if the problems can be overcome,
then a muon-muon collider may be a viable tool for the study of high energy
phenomena, complementary to \ee and hadron colliders.
 \end{itemize}
\section{ACKNOWLEDGMENTS}
We acknowledge extremely important contributions from our colleagues,
especially W. Barletta, A. Chao, D. Cline, D. Douglas, D. Helms, J.
Irwin, K. Oide, H. Padamsee, Z. Parsa, C. Pellegrini, A. Ruggiero,  W. Willis
and Y. Zhao.

\medskip
This research was supported by the U.S. Department of Energy under Contract No.
DE-ACO2-76-CH00016 and DE-AC03-76SF00515.

\end{document}